\documentclass[os,manuscript]{copernicus}  

\begin{document}  
\nolinenumbers  

\title{A Comparative Analysis of Clustering Algorithms for Characterizing Surface Ocean Variability in the Western Mediterranean}

\Author[1]{Victor}{Rodr\'{\i}guez-M\'endez}
\Author[1]{Enrico}{Ser-Giacomi}
\Author[1]{Jos\'e J.}{ Ramasco}
\Author[1]{Crist\'obal}{L\'opez}
\Author[1][emilio@ifisc.uib-csic.es]{Emilio}{Hern\'andez-Garc\'{\i}a}

\affil[1]{Instituto de F\'{\i}sica Interdisciplinar y Sistemas Complejos, IFISC (UIB-CSIC), Campus Universitat de les Illes Balears, 07122 Palma de Mallorca, Spain}

\runningtitle{Comparing Clusters of Ocean Variability in the Western Mediterranean}

\runningauthor{V. Rodr\'{\i}guez-M\'endez et al.}

\received{May 13, 2026}
\pubdiscuss{} 
\revised{}
\accepted{}
\published{}


\firstpage{1}

\maketitle

\begin{abstract}
Understanding regional dynamical structures in the sea is fundamental to characterize energy transfer and transport properties, with implications in physical and biogeochemical modeling and characterization. In this work, we study the potential of clustering techniques to identify regional patterns, persistent or recurrent configurations, out of daily snapshots of sea surface temperature and kinetic energy in a region of the western Mediterranean Sea. From the methodological perspective, we use different clustering techniques: K-means,  Self-Organizing Maps and InfoMap to verify if the patterns found are coherent across methods. Our results show that K-means and Self-Organizing Maps consistently  delineate four distinct clusters of sea surface temperature configurations, aligned with the seasons even after removing the annual cycle, which indicates the persistence of seasonal structures beyond a mean effect in the temperature field. The study of surface kinetic energy, characterized by higher spatial and temporal variability, reveals more complex circulation regimes. While K-means and Self-Organizing Maps provide a robust and convergent classification of the dominant large-scale energy patterns, InfoMap uncovers finer-scale features such as localized jets and eddies. InfoMap, in particular, provides a complementary perspective to the partition-based methods, validating subtle yet significant hydrodynamic structures and acting as an anomaly detector for extreme events.
\\
\textbf{Date:} May 13, 2026
\end{abstract}

\introduction  

The ocean plays a crucial role in Earth's climate system, and understanding its complex dynamics is essential
for environmental research and management. Analyzing temporal snapshots of oceanic variables such as kinetic
energy (KE) and sea surface temperature (SST), provides valuable insights into the spatiotemporal
structures and processes occurring within the marine environment~\citep{Liu2006, Richardson2003}. The increasing
availability of extensive and synoptic oceanographic datasets has promoted the development of advanced analysis techniques,
as discussed by \citet{thomson2014data} and \citet{Halkidi2001}. Cluster analysis, a fundamental tool in data mining and pattern recognition, has
emerged as a powerful approach for identifying groups of similar configurations and discovering distribution trends
in large datasets \citep{Zait1997}.

Recently, two clustering algorithms that have gained prominence in oceanographic research are K-means and Self-Organizing Maps (SOM) \citep{Liu2016,HernandezOrfila2018,Hernandez2018,Finley2024}.  K-means is a widely used partitioning algorithm introduced by \citet{MacQueen1967}, which has been successfully applied in various environmental studies, from precipitation data \citep{Gong1995} to classifying water mass types based on temperature and salinity \citep{Emery1986global}. Its simplicity and efficiency make it an attractive option for clustering large datasets, although it can be sensitive to initial conditions \citep{Celebi2013}.

SOM, an unsupervised neural network technique introduced by \citet{Kohonen1982}, has proven particularly effective in extracting structures from satellite imagery and other oceanographic observations \citep{Richardson2003, Liu2005}. Its ability to project high-dimensional data onto a lower-dimensional grid while preserving topological relationships makes it well-suited for visualizing and analyzing complex oceanographic phenomena. \citet{Vesanto2000} expanded the analytical scope of SOM by introducing a two-stage clustering approach, a methodology that has proven highly effective for identifying the characteristic signatures of complex climate phenomena \citep{Leloup2007}.
Recent studies have demonstrated the power of SOM for analyzing oceanic dynamics. \citet{HernandezOrfila2018} used a combined space- and time-domain SOM analysis of altimetry-derived finite-size Lyapunov exponents to classify the western Mediterranean into regions of different mixing-activity variability. Their work revealed a previously unnoticed recurrent and intense front between Cartagena, Spain, and Tenes, Algeria, showing the potential of SOM for uncovering important mesoscale oceanic structures.

The integration of different clustering techniques has proven beneficial in oceanographic research. \citet{Fernando2005} explored the use of SOM  as substitutes for K-means clustering. Building on this idea, \citet{Solidoro2007} demonstrated the power of combining SOM with K-means clustering to analyze biogeochemical properties in coastal waters. This hybrid approach reduces the complexity of spatiotemporal variations in multiple variables to a single categorical variable representing water typology.

Thus, while both SOM and K-means have been used in oceanographic research, there is a growing need for comparative studies that evaluate the performance gained from these different clustering approaches \citep{Budayan2009, Kumar2010}. Such comparisons, which is one of the main objectives of the present paper, can help researchers select the most appropriate method for their specific data and research questions, as well as provide more understanding of the underlying oceanic patterns.

Another clustering method that uses rather different principles is Infomap \citep{Rosvall2008}. It is a network-based approach that uses the probability flow of random walks on a network as a proxy for information flows in the real system. Finding the minimum-information necessary to label the different walks determines clusters of nodes in the network which share relevant properties. InfoMap's effectiveness has been demonstrated in various fields, including neuroscience \citep{Power2011} and ecology \citep{Lancichinetti2012}. Oceanographic applications include the identification of connected regions in the Mediterranean Sea based on surface flow dynamics \citep{Ser-Giacomi2015}. However, it is important to note that InfoMap is more computationally costly than K-means or SOM. 

Our study applies and compares SOM, K-means, and InfoMap clustering techniques in oceanographic data from the surface of the Western Mediterranean. In particular, we analyze the spatial distribution of SST and surface KE per unit mass and their temporal variability in the surface area around the Ibiza channel, Spain, i.e. between the Balearic islands of Ibiza and Formentera and the Spanish coast. This region is selected because of its importance as meeting point between the Atlantic waters entering through Gibraltar and the saltier waters than form across the Mediterranean basin, with the presence in the zone of active mesoscale structures such as fronts and eddies. Our objectives are twofold. First, we aim at identifying meaningful patterns within these oceanic variables. Second, we want to compare the results obtained from the three clustering methods, evaluating their respective strengths and limitations.

The outline of the paper is the following:  Next section describes the data  and the different clustering methodologies used in this work. Section \ref{sec:results} describes first the silhouette analysis performed, and then applies the clustering methodologies to SST (Sect. \ref{subsec:resultsSST}) and to KE (Sect. \ref{subsec:resultsKE}). Section \ref{subsec:resultsjaccard} compares the cluster partition obtained by the different methods, and our Conclusions are summarized in Sect. \ref{sec:conclusions}.

\section{Dataset and Methods}
\label{sec:datamethods}
Here we describe the dataset and the
methods used to analyze them (K-means, SOM and infoMap).
We also describe two metrics to compare the obtained clusters,
silhouette score and Jaccard similarity. Our Python code implementing these operations, as well as the datasets analyzed, are available from \citet{DataCode}. 

\subsection{Data Description and Preprocessing}
\label{subsec:data}

This study utilizes the Mediterranean Sea Physical Reanalysis Product (MEDSEA\_MULTIYEAR\_PHY\_006\_004), supplied by the Copernicus Marine Environmental Monitoring Service (CMEMS). The reanalysis system is described in detail by \citet{Escudier2021}.

\begin{figure}[t]
    \includegraphics[width=8.3cm]{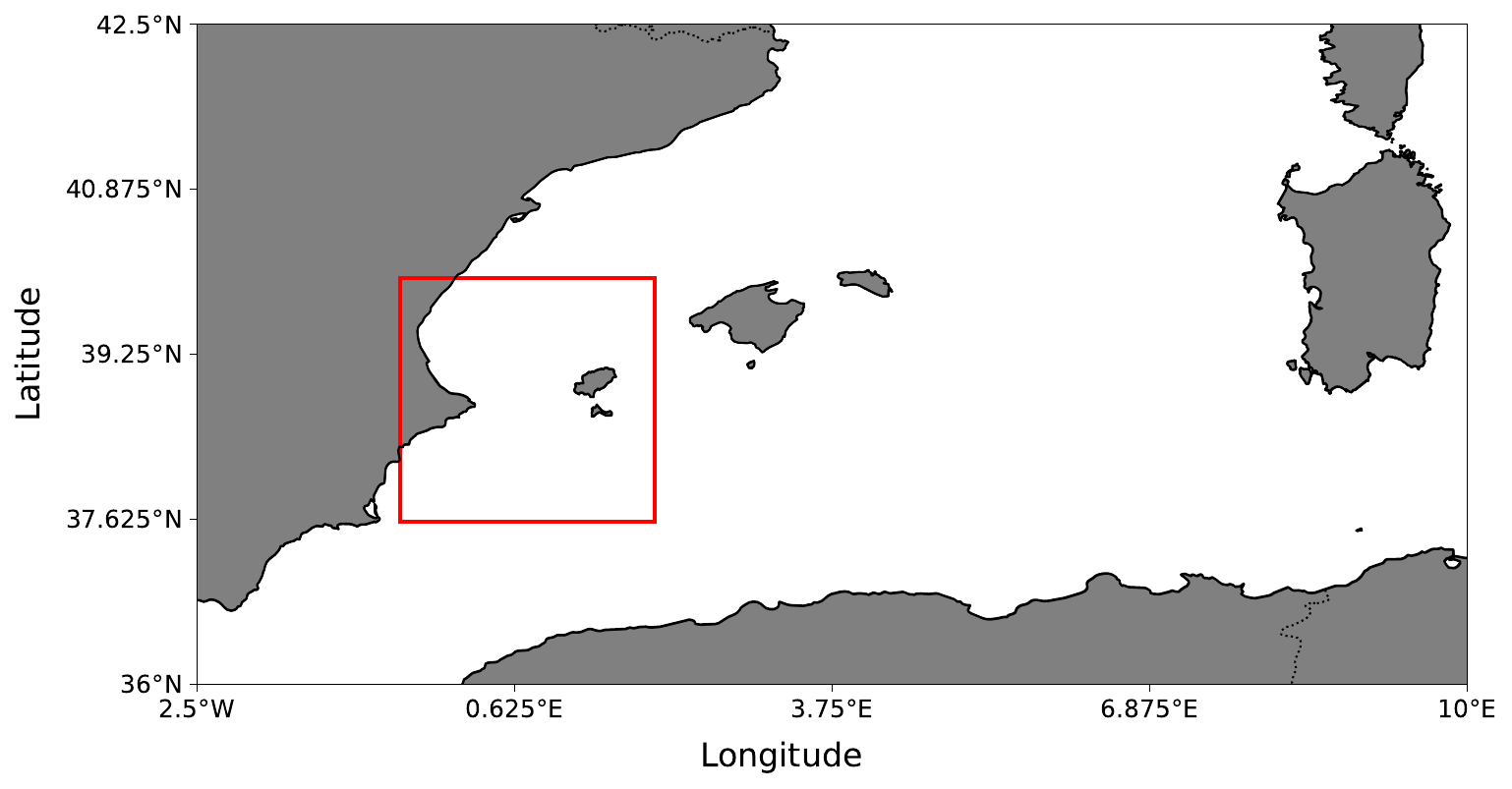}
    \caption{The study region (red box) within the broader context of the Western Mediterranean Sea.}
    \label{fig:region}
\end{figure}

The MEDSEA reanalysis provides high-resolution physical ocean data for the Mediterranean Sea by integrating satellite and in-situ observations with the NEMO (Nucleus for European Modelling of the Ocean) ocean model. This model is coupled with a variational data assimilation scheme (OceanVAR), a statistical method for combining observations and model predictions to minimize errors \citep{Escudier2021}.
We extract from this dataset the SST data
(uppermost layer temperature) and the surface zonal ($u_0$) and meridional ($v_0$)
velocity components, from which we compute the surface KE per unit mass as  $KE = (v_0^2 + u_0^2)/2$ (the energetic contribution of the vertical velocity component is completely negligible). We consider the temporal interval from January 1, 1987 to July 31,
2022, at daily resolution, which gives us $N=12996$ snapshots (instantaneous spatial configurations) of SST and of KE.
The study region will be the one limited by the latitudes
37.6$^\circ$N to 40$^\circ$N, and the longitudes 0.5$^\circ$W to 2$^\circ$E (see Fig.
\ref{fig:region}). This correspond to the area of the Ibiza Channel, between the Mediterranean Spanish coast and
the Island of Ibiza, in the Western Mediterranean.
Data spatial resolution is  $\frac{1}{24}^\circ$ (approximately 4-5 km), so that each snapshot is an image with $M=3136$ pixels for each variable (taking only the sea pixels, discarding the ones over land). Spatiotemporal data are conveniently arranged into $N\times M$ reshaped feature matrices $X_{t, s}^{(k)}$. $t = 1, 2, ..., N$ is the time index and $s = 1, 2, ..., M$ is the spatial index. $k=1,2$ indicates the specific variable (SST or KE, respectively).

\subsection{K-means clustering}
\label{subsec:kmeans}

K-means clustering \citep{MacQueen1967} is an unsupervised
learning algorithm that partitions a dataset into $K$ distinct,
non-overlapping subgroups or clusters. The
algorithm aims to minimize the within-cluster sum of squares (WCSS):
\begin{equation}
    WCSS = \sum_{k=1}^{K} \sum_{\mathbf{x} \in C_k} \|\mathbf{x} - \mathbf{\mu_k} \|^2,
\end{equation}
where each $\mathbf{x}$ is a single temporal snapshot of the study area, constructed as an $M$-dimensional feature vector. Each component of $\mathbf{x}$ corresponds to the pixel intensity (the specific value of SST or KE) at a given spatial location $s \in \{1, ..., M\}$, effectively reshaping the 2D image into a 1D vector to characterize the spatial configuration of the field at a specific time $t$. In total there are $N$ different snapshots. $K$ is the number of clusters, $C_k$ is the set containing the elements of the $k$-th cluster. $\mathbf{\mu_k}$ is the centroid of cluster $C_{k}$, defined as:
\begin{equation}
    \mathbf{\mu_k} = \frac{1}{|C_k|} \sum_{\mathbf{x} \in C_k} \mathbf{x} \ ,
\end{equation}
where $|C_{k}|$ is the number of elements in that cluster. The algorithm requires the number of clusters $K$ as an input and produces the clusters and their corresponding centroids as output. The basic K-means algorithm proceeds as follows:
\begin{enumerate}
    \item Initialize $K$ cluster centroids randomly.
    \item Assign each snapshot $\mathbf{x}$ to the nearest centroid, grouping the $N$ data points into $K$ clusters.
    \item Recalculate the centroids as the mean of all data points in each cluster.
    \item Repeat steps 2 and 3 until convergence or maximum iterations are reached.
\end{enumerate}
Due to its sensitivity to initial centroid positions, the algorithm was run multiple times with different random initializations, and the best result (lowest WCSS) was retained.

\subsection{Self-Organizing Maps}
\label{subsec:som}

SOM \citep{Kohonen1982} is an unsupervised learning algorithm that produces a low-dimensional, typically two-dimensional, representation of high-dimensional input data. The basic architecture of SOM is
a grid of elements, called \textit{neurons}, that compete to represent different
regions of the input data space, thus defining the different data clusters. Each neuron has an associated weight vector of $M$ components
(called the \textit{codebook vector}) that is representative of data in its cluster. During the learning process, neurons become selectively responsive to various input patterns, and the SOM organizes itself topologically, such that similar input patterns activate neighboring neurons on the grid. This results in a 'map' that maintains the topological structure among data points, supporting the visualization and interpretation of high-dimensional datasets.

The input for SOM, like for K-means, are the $N$ snapshots of one of the physical fields, obtained at different times. However, while K-means treats all snapshots simultaneously, the codebook vectors of the SOM neurons reach their final state, representing learned spatial patterns, by sequentially processing the snapshots from oldest to newest. In addition to the number of clusters, the topology of the grid of neurons (i.e.
which are the neighbors of each neuron in the grid) should be specified as input. The outputs of the SOM algorithm are data-point clusters, each represented by a SOM neuron, and the codebook or weight vector associated to each of them, representative of the snapshots or spatial configurations assigned to that cluster. The SOM algorithm can be summarized as follows:
\begin{enumerate}
    \item Initialize the weight vectors $\mathbf{w}_i$ for each neuron $i$ randomly. Each weight vector $\mathbf{w}_i$ has the
    same dimensionality $M$ as the input data.
    \item For each input data point, $\mathbf{x}$ (processed in temporal order):
    \begin{enumerate}
        \item Find the Best Matching Unit (BMU), denoted by $c$:
        \begin{equation}
            c = \underset{i}{\operatorname{arg\,min}} \|\mathbf{x} - \mathbf{w}_i\|.
        \end{equation}
        Thus, the BMU is the neuron whose weight vector $\mathbf{w}_c$ is most similar to the input data point $\mathbf{x}$,
        measured by the Euclidean distance.
        \item Update the weight vectors of the BMU and of its neighbors (i.e., in the following formula
        the neuron index $i$ takes the values $c$ and the values corresponding to $c$'s neighbors in the SOM grid):
        \begin{equation}
            \mathbf{w}_i(t+1) = \mathbf{w}_i(t) + \alpha(t) h_{ci}(t) [\mathbf{x}(t) - \mathbf{w}_i(t)] \ ,
        \end{equation}
        where:
        \begin{itemize}
            \item $\mathbf{x}(t)$ is the input data point at time $t$ (a spatial configuration or snapshot of either SST or KE).
            \item $\mathbf{w}_i(t)$ is the weight vector of neuron $i$ at time $t$.
            \item $\mathbf{w}_i(t+1)$ is the updated weight vector of neuron $i$ at time $t+1$.
            \item $\alpha(t)$ is the learning rate at time $t$, a scalar value between 0 and 1 that controls the size of
            the weight updates at each step. It is taken to decrease  monotonically with time.
            \item $h_{ci}(t)$ is the neighborhood function at time $t$, which defines the influence of the input data point
            $\mathbf{x}(t)$ on the weight vectors of neurons $i$ in the neighborhood of the BMU $c$. It is a function that decreases
            with the distance between neuron $i$ and the BMU $c$ on the SOM grid, and also decreases with time.
        \end{itemize}
    \end{enumerate}
 \item Repeat step 2 for a pre-defined number of epochs (iterations).
\end{enumerate}


The SOM training was performed using the complete dataset of $N$ daily snapshots.
To differentiate between training steps and data quantity, we define the training duration in terms of epochs, where one epoch represents one full pass through the entire dataset. The model was trained for 50 epochs, resulting in a total number of iterations $T = 50 \times N$.
During this process, the weight vectors were updated sequentially. The learning rate $\alpha(t)$ follows a decay function,
$$\alpha(t) = \frac{\alpha_{0}}{1 + \frac{2t}{T}}$$
where $\alpha_{0} = 0.1$ is the initial learning rate and $t$ is the current iteration step ($0 \le t \le T$).
The topological preservation is achieved via the neighborhood function $h_{ci}(t)$, which determines the extent to which the BMU ($c$)
influences its neighbors. We utilize a Gaussian neighborhood function based on the Euclidean distance between the grid coordinates of the BMU ($\mathbf{r}_c$) and a neighbor neuron $i$ ($\mathbf{r}_i$):
$$h_{ci}(t) = \exp \left( - \frac{\| \mathbf{r}_c - \mathbf{r}_i \|^2}{2\sigma(t)^2} \right)$$.
The neighborhood radius $\sigma(t)$, which controls the width of this Gaussian, decays asymptotically from an initial value of $\sigma_{0} = 1$ (effectively covering immediate neighbors) according to:
$$\sigma(t) = \frac{\sigma_{0}}{1 + \frac{2t}{T}}\ .$$
The chosen parameters enable  SOM to effectively learn and organize the high-dimensional temporal snapshots onto a lower-dimensional grid, revealing underlying spatiotemporal relationships in the oceanographic observations by grouping similar configurations together

\subsection{InfoMap Clustering}
\label{subsec:infomap}

To further explore the complex hydrodynamic patterns in our study area, we applied the InfoMap community detection method to the KE field. InfoMap \citep{Rosvall2008} is a network-based clustering algorithm that leverages the probability flow of random walks to reveal community structures. It identifies partitions where information flow is most confined, thus uncovering inherent communities. The method operates by minimizing the expected description
  length of a random walk through the network.

In our application to the surface KE per unit mass field, we constructed networks from our data, where each daily configuration or snapshot of KE represents a node. To define the connections (links) between these nodes, we calculated the cosine similarity between all pairs of snapshots. This means that higher cosine similarity values (indicating greater similarity in KE spatial distribution regardless of the absolute KE values) resulted in stronger link weights. To focus on these stronger relationships and filter out weaker noisy connections, we retained only those links whose weight was greater than a given threshold. This threshold was set at the mean cosinus similarity plus 1.27 times the standard deviation of all similarity values. Thus, this approach emphasizes the most significant spatial pattern similarities in the KE field, translating them into network connections. Once the network is defined the algorithm is run (see details in \citet{Rosvall2008}) to find a classification of the configurations
into separate communities or clusters. The method provides the optimal number of clusters found in the data, although
 this greatly depends on the choice of the similarity threshold used to define the link. It can be thought that the number of
clusters is indirectly set by the choice of the threshold, since the threshold determines how many similarity links are retained in the weighted network.

\subsection{Silhouette score}
\label{subsec:silhouette}

Once we have the clustering results given by any of the
previous methods, we use the silhouette score
to evaluate the quality of these clustering results. It measures how similar an object
is to its own cluster (cohesion) compared to other clusters (separation).
Most importantly, and this is the way we will use it, the silhouette analyses can be used to determine
an optimal number of clusters for K-means and SOM.
The silhouette score for a single data point {\it i} is calculated as:

\begin{equation}
s(i) = \frac{b(i) - a(i)}{\max{(a(i), b(i))}},
\end{equation}
where {\it a(i)} is the mean distance between data point {\it i}
and all other points in the same cluster, and {\it b(i)} is the minimum mean distance
 between data point {\it i} and all other points in a different cluster, where the minimum is taken over these different clusters. We have $s(i) \in [-1,1]$, with values close to 1 indicating that point $i$
has been well classified in a given cluster, whereas null or negative values indicate that this point has been poorly classified. Standard practice selects $k$ by maximizing the mean silhouette value $\overline{s}(k)$ (mean over data points $i$), but we instead identify the 'knee' or break point of the curve $\overline{s}(k)$ aa a function of $k$ (see Sect. \ref{subsec:resultssilhouette}). This heuristics finds the most complex natural structure before clusters begin to fragment artificially. By following this approach, we ensure a sufficiently rich set of representative patterns while maintaining high cluster quality \citep{rousseeuw1987silhouettes}.

\subsection{Jaccard Similarity}
\label{subsec:jaccard}

Another useful metrics to analyze the results from the different clustering methods is the Jaccard similarity index.
This quantitatively assesses the concordance
between the clusters obtained from one method or from another,
It is defined as the ratio between the sizes of the intersection and the union of the data points in two clusters,
\begin{equation}
J(C^A_{m}, C^B_{n}) = \frac{|C^A_{m} \cap C^B_{n}|}{|C^A_{m} \cup C^B_{n}|},
\end{equation}
Here, $C^X_j$ is cluster $j$ obtained through clustering method $X$, and $|C|$ is again the number of configurations in cluster $C$. $J(C^A_{m}, C^B_{n})  \in [0,1]$, with values close to 1 indicating
that the two clusters contain nearly the same elements and values close to zero indicating nearly disjoint clusters.
The matrix $J_{mn}\equiv J(C^A_{m}, C^B_{n})$ allows to identify
which cluster obtained from method $A$ is more similar to a given cluster obtained from method $B$, and how consistent are the two clustering
methods among them.

\section{Results}
\label{sec:results}

We have applied K-means and SOM to SST and KE data for several values of the a-priori-fixed number of clusters and their topology.
To select appropriate values of these parameters, we first explore several of them and evaluate their mean silhouette scores. This is described in the following section. Later in this Results section we will also consider detrended SST data and the use of InfoMap for KE.





\subsection{Silhouette score analysis}
\label{subsec:resultssilhouette}

\begin{figure}[t]
    \includegraphics[width=8.3cm]{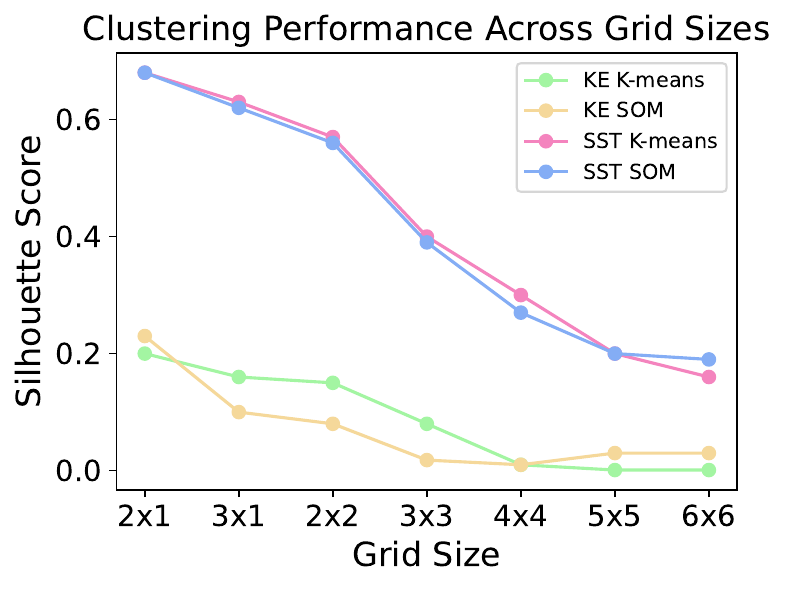}
    \caption{Mean silhouette scores for different clustering methods. Scores for KE and SST using K-means and SOM across varying grid sizes (number of clusters) and topologies. The significant drop in mean silhouette score around four clusters for both K-means and SOM (with $2\times2$ topology in this case), particularly pronounced for SST, supports the selection of four clusters for subsequent analyses.}
    \label{fig:grid_size_vs_silhouette}
\end{figure}

The mean silhouette is a guide to locate an optimal number of clusters to be used in K-means and SOM clustering.
In Fig. \ref{fig:grid_size_vs_silhouette} we show this analysis (for both methods and for both SST and KE datasets).
The mean silhouette score is plotted for each predetermined number of
clusters. For the case of SOM, not only the number of clusters is relevant, but also its topology. Thus we indicate in the horizontal axis this topology. For example, data for grid size $3\times 3$ means that we use 9 clusters in both K-means and SOM, and in this last method they  are arranged in a square lattice with three rows and three columns of clusters.
For SST, we observe that the silhouette score exhibits a sharp decline around a number of four clusters.
This suggests that considering four clusters
provides a well-balanced partitioning of the SST configurations, including a sufficiently diverse set of clusters, but still with
a significant score before the quality loss indicated by the silhouette drop. The same is observed for KE, although in this case scores are lower, indicating a poorer identification of distinct clusters. In the following we will fix the number of clusters to four ($2\times 2$) in the clustering analyses.

\begin{figure}[t]
    \centering
    \includegraphics[width=8.3cm]{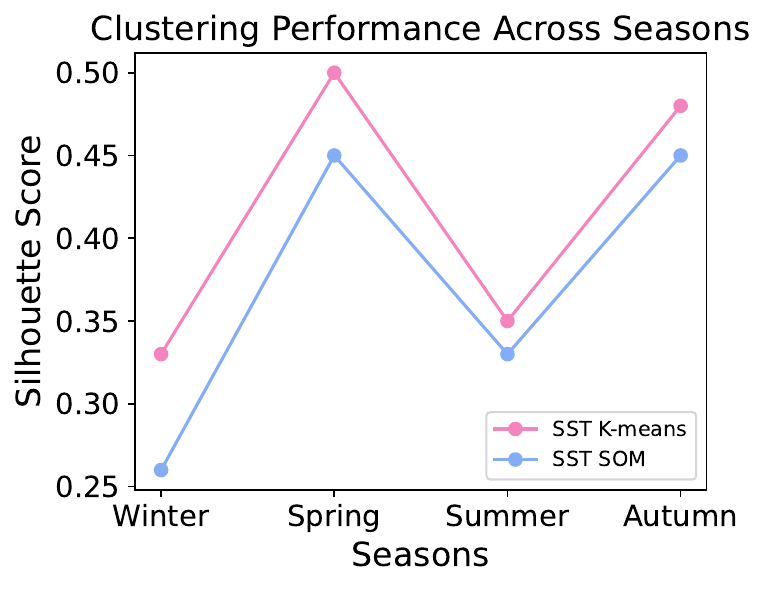}
    \caption{Mean silhouette scores for SST clustering across meteorological seasons (4-clusters configuration). }
    \label{fig:seasons_vs_silhouette}
\end{figure}

In Fig. \ref{fig:seasons_vs_silhouette} we compare the clustering quality for SST data when using only configurations from each of the four seasons (both for K-means and SOM). Seasons are here conventionally defined so that winter configurations are those occurring between December 21 of each year until March 20, spring from March 21 until June 20, etc.
We found an interesting pattern: the scores for winter and summer were always lower than the scores for spring and autumn.
This happens because the variability of SST patterns during the winter and summer months is smaller than in the other seasons, and forcing them to be classified in four different clusters leads to poor separation and then lower silhouette. In contrast, during the transitional seasons of spring and autumn, SST configurations fluctuate more, facilitating a consistent classification of them in four clusters. This seasonal pattern makes sense for the Mediterranean climate, where winter and summer are known for having more stable and consistent temperatures.

We note in Figs. \ref{fig:grid_size_vs_silhouette} and \ref{fig:seasons_vs_silhouette} the
great similarity for both the SOM and the K-means analysis.
This similarity suggests that both methods are equally capable of uncovering the underlying structure within
the data. Whether it is the grid-based organization of SOM or the centroid-driven approach of K-means, both methods yielded comparable clustering quality, reinforcing mutually the validity of both methodologies. In the following we analyze more in detail the nature of the
clusters of configurations found.

\subsection{Sea Surface Temperature}
\label{subsec:resultsSST}

\subsubsection{K-means for SST}

\begin{figure}[t]
    \centering
    \includegraphics[width=0.5\textwidth]{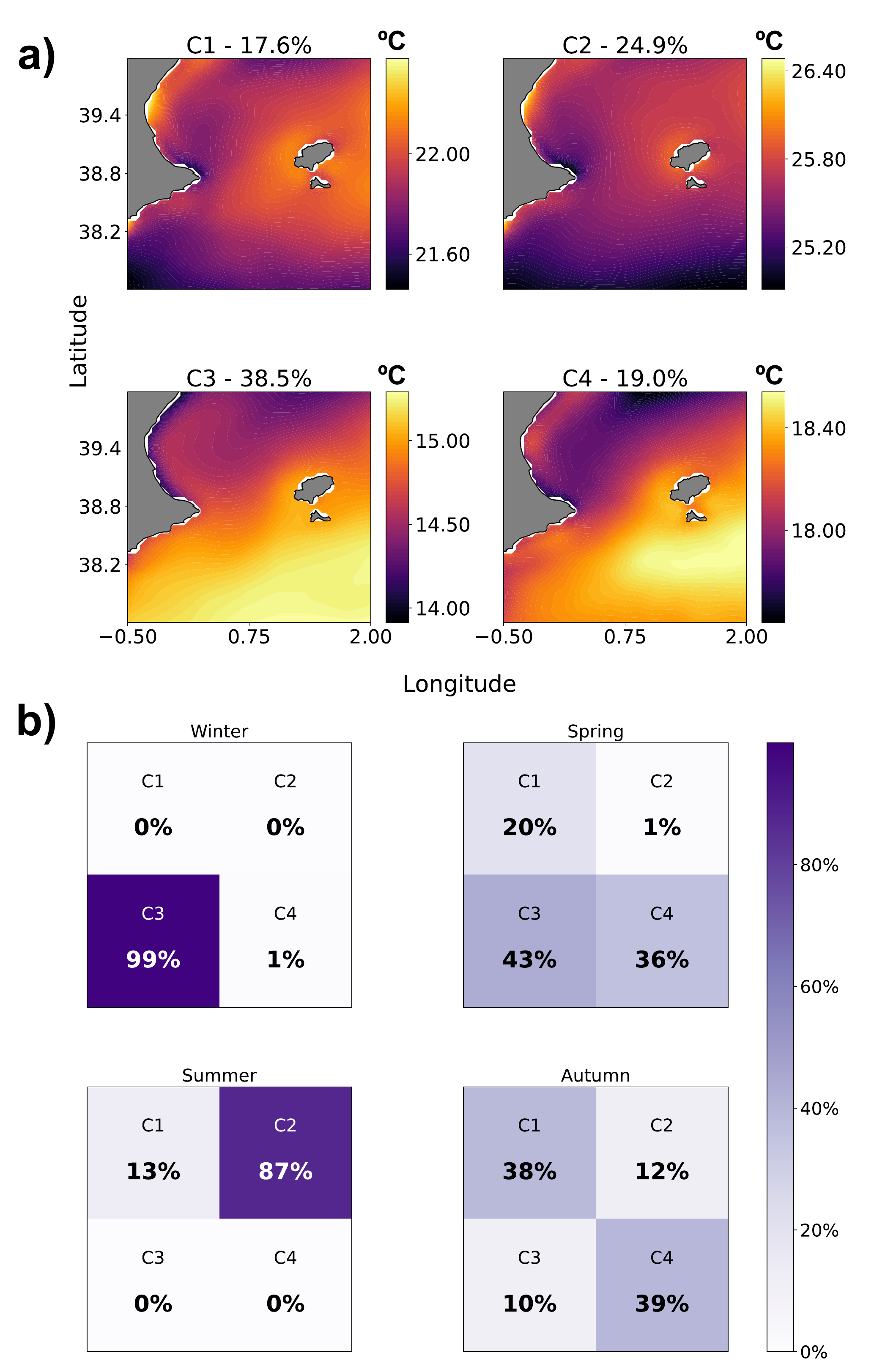}
    \caption{{\bf K-means Clustering of Sea Surface Temperature Data.}
(a) Cluster Centroids: Spatial distribution of the four cluster centroids (C1-C4) derived from K-means clustering of SST data. The percentage of data points assigned to each cluster is indicated.
(b) Seasonal Distribution of SST configurations in clusters: Proportion of observations within each cluster across the four seasons (Winter, Spring, Summer, Autumn). This visualization illustrates the seasonal prevalence of each cluster, revealing the strong relationship between SST clusters and the seasonal cycle. }
    \label{fig:seasons_and_clusters_kmeans_temp}
\end{figure}

In this section we apply the K-means algorithm
with a predetermined number of 4 clusters, as
suggested by the silhouette analysis.
Figure \ref{fig:seasons_and_clusters_kmeans_temp}a shows the centroids of the SST spatial patterns in each cluster. So far, the labelling of the clusters as C1, C2, C3, and C4 is arbitrary, but we are already choosing the names that will result in the best similarity (see Sect. \ref{subsec:resultsjaccard}) with the corresponding clusters obtained with the SOM method, for which the labelling has a topological significance.

The first observation is that, in the ocean area of study, the mean SST of the centroid patterns are quite different. They are ordered, from low to high temperatures, as C3, C4, C1, C2. Thus C3 is associated to low SST values, mostly occurring in winter, and C2 to high temperatures, more likely to occur in summer. Clusters C1 and C4 would collect, respectively, hot and cold subsets of transitional states. It should be mentioned that a significant proportion of spring configurations are also assigned to C3, so that configurations in this cluster are the most frequent along the year.

The alignment of the clusters with the seasonal cycle is further supported by Figure \ref{fig:seasons_and_clusters_kmeans_temp} b). It presents the seasonal distribution of data points within each cluster. As anticipated, C3 exhibits a strong dominance of winter configurations, with 99\% of winter entries. This result aligns with expectations, since winter SST fields typically exhibit greater spatial and temporal homogeneity compared to the stratified seasons \citep{DOrtenzio2005}. Also, C2 configurations are predominant in summer.

In general, the centroid SST configurations are smooth and relatively featureless, as corresponding to their average nature. The winter-associated C3 centroid shows a clear northwest-southeast temperature gradient, whereas the summer-associated C2 is more homogeneous, with somehow lower temperature in the southern border of the region (also present in C1), which we interpret as a signal of Atlantic waters. In all patterns SST increases close to the islands, and there is a hotter tongue appearing from the west in C4.


\begin{figure*}[t]
    \includegraphics[width=12cm]{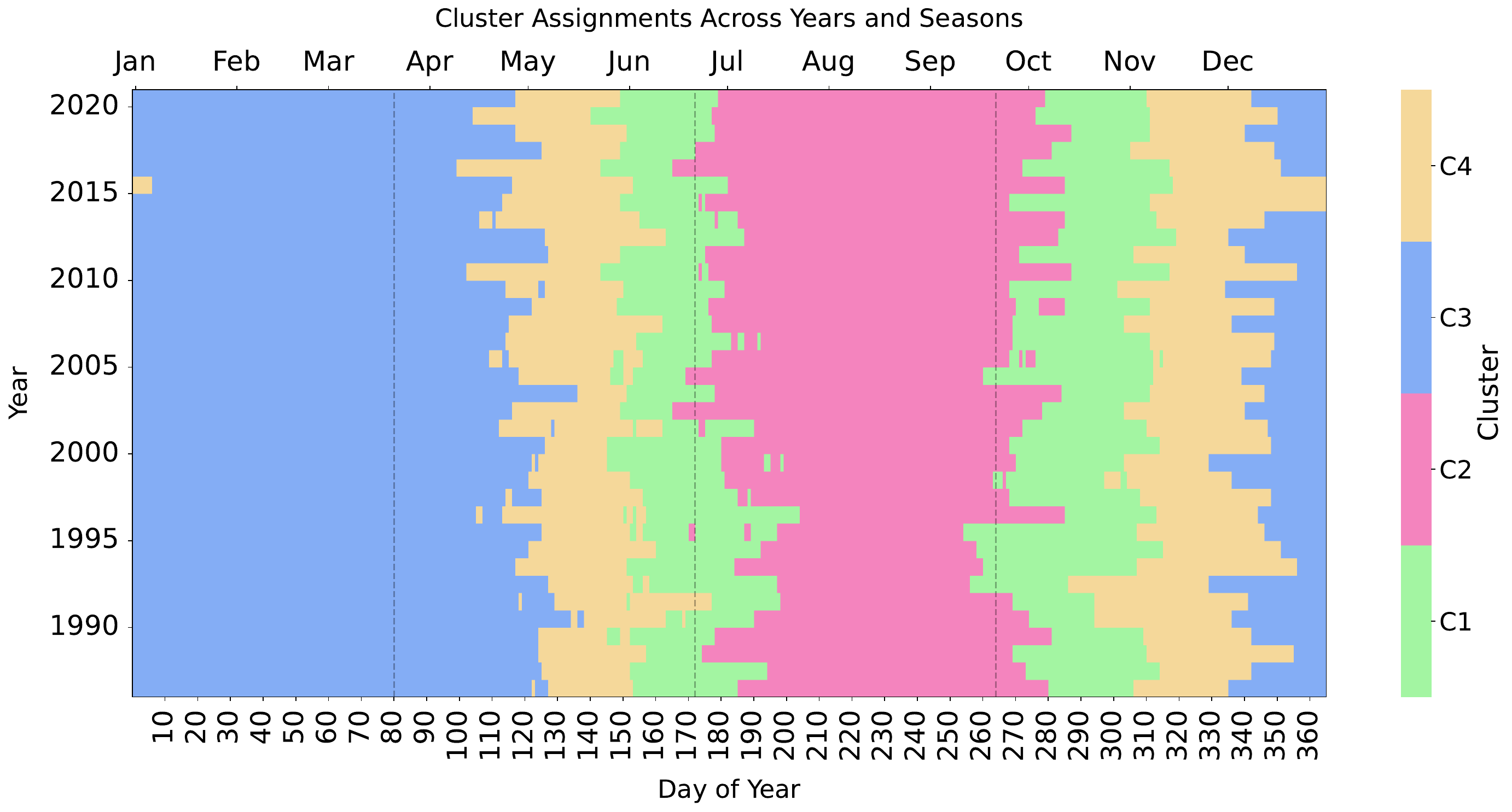}
    \caption{Temporal distribution, from 1987 to 2022, of SST states in the four clusters (C1–C4) identified through K-means clustering. Vertical dashed lines mark conventional seasonal boundaries (21 December, 21 March, 21 June, and 21 September). Each color represents a different SST cluster, demonstrating both seasonal patterns and inter-annual variability.}
    \label{fig:cluster_heatmap_thetao_kmeans}
\end{figure*}

In Fig. \ref{fig:cluster_heatmap_thetao_kmeans},
we show the cluster distribution along the 35-year study period (1987-2022).
There is a clear banding in the figure signalling the strong
correspondence with the seasons, so that each cluster's
dominance aligns with specific months of the year,
consistently across the study period. This confirms the character of C3 and C2 as associated to winter and summer, and the transitions between these regimes through the C4-C1 sequence (in spring, linking winter to summer) and the C1-C4 sequence (in autumn, linking summer to winter). Beyond the broad seasonal
alignment, the figure also captures
inter-annual variations in the timing
and duration of seasonal transitions.

The figure reveals also a notable trend towards earlier
 termination of configurations in the winter-associated C3 cluster in recent years compared
 to the beginning of the study period, visible as a leftward shift in the
 C3-C4 boundary from the bottom to the top of the
 figure. This suggests a trend towards shorter
  winters in the Mediterranean region, reflecting
  shifts in the timing and duration of seasonal transitions
\citep{giorgi2008climate}. An associated increase in the duration of the states in the summer-associated C2 cluster is also seen.

\subsubsection{K-means for detrended SST}

\begin{figure}[t]
    \centering
    \includegraphics[width=0.5\textwidth]{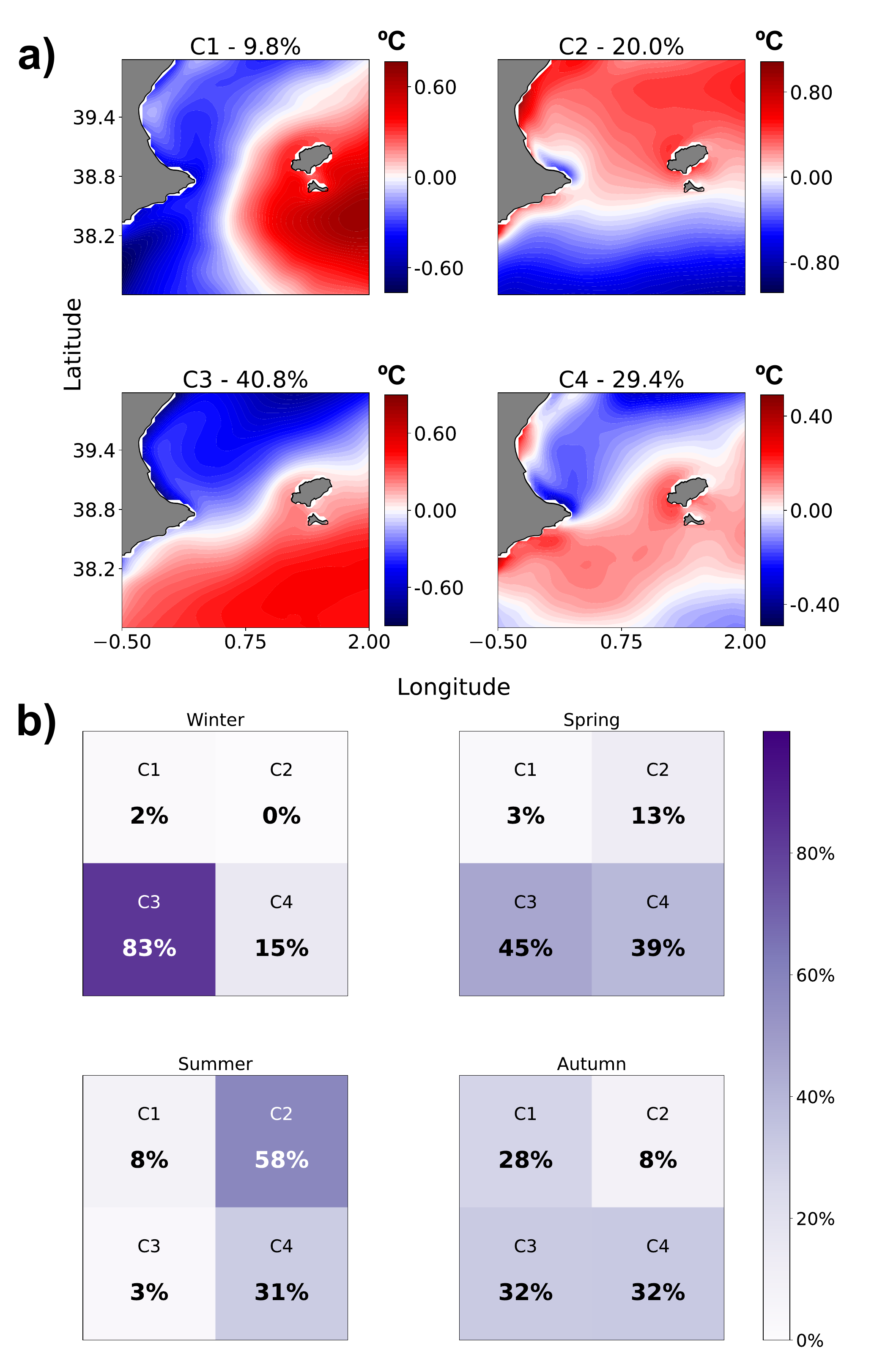}
    \caption{{\bf K-means Clustering of Spatially detrended SST data.} (a) Cluster Centroids: Spatial patterns of the four centroids (C1–C4) obtained from K-means clustering applied to SST data with the spatial mean removed from each daily snapshot. The percentage of data points assigned to each cluster is indicated. (b) Seasonal Distribution of Clusters: Proportion of spatially detrended SST observations within each cluster across the four seasons (Winter, Spring, Summer, Autumn).}
    \label{fig:seasons_and_clusters_kmeans_temp_subtracted}
\end{figure}
\begin{figure*}[t]
    \centering
    \includegraphics[width=12cm]{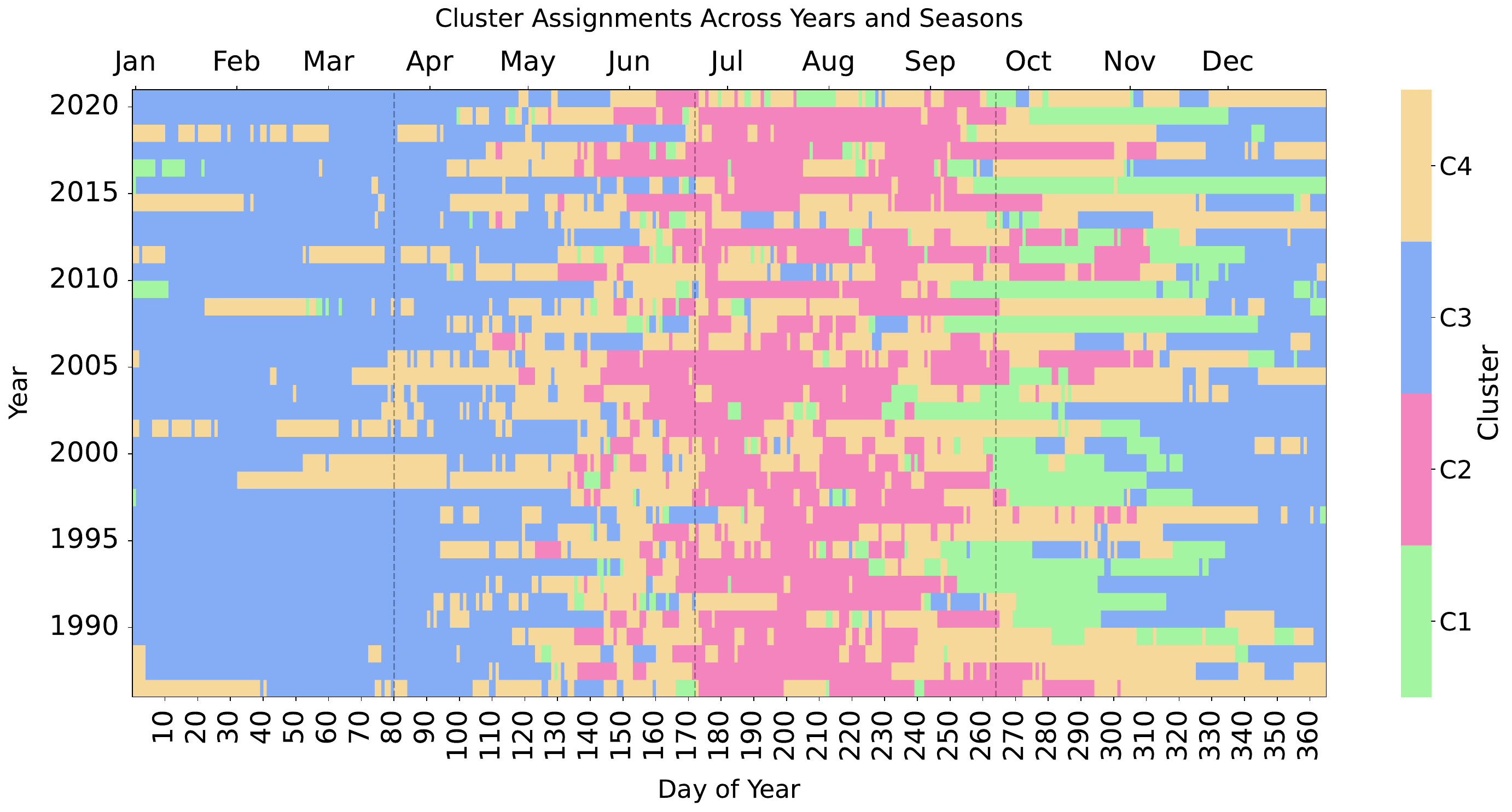}
    \caption{Temporal Distribution of detrended SST clusters in the Mediterranean Sea from 1987 to 2022. The plot shows daily cluster assignments (C1–C4) resulting from K-means clustering of spatially detrended SST data. Vertical dashed lines mark conventional seasonal limits. Cluster colors represent the identified distinct SST-anomaly regimes, revealing recurrent seasonal transitions and the persistence of sub-seasonal spatial variability over multiple decades.}
    \label{fig:cluster_heatmap_thetao_kmeans_subtracted}
\end{figure*}

The above results confirm the appropriateness of the choice of four clusters to describe the SST dynamics of our area of study, as being clearly linked to the seasonal cycle. But it also questions the significance of the spatial details of the centroids observed in Fig \ref{fig:seasons_and_clusters_kmeans_temp}, as it may happen that the only feature in the data set that the clustering method is detecting is the mean temperature in the region.  To check whether or not this is the case,
we apply now the K-means algorithm to SST data
for which the spatial mean has been subtracted from
each daily snapshot. With this we aim at removing
the dominant large-scale seasonal SST signal. We use again a predefined number of 4 clusters.
The resulting cluster centroids, presented
in Figure~\ref{fig:seasons_and_clusters_kmeans_temp_subtracted} a),
 show coherent patterns of temperature anomalies
 that capture intrinsic variability within the region.
The seasonal distribution of cluster assignments is shown in
Fig.~\ref{fig:seasons_and_clusters_kmeans_temp_subtracted}, panel b), and the cluster distribution over months and the 35 years of study is in Fig. \ref{fig:cluster_heatmap_thetao_kmeans_subtracted}. The figures show that despite the much noisier cluster assignment after the detrending, still clusters C3 and C2 are associated to winter (83\% of wintertime days) and summer (58\% of summer days), respectively. The centroid of C3 shows the pattern observed before of relatively hotter temperatures in the southeast, and colder in the northwest, and the centroid of C2 still displays the colder waters in the southern part. The transitional state C1 is now less represented, occurring only during some years during autumn, and all the spring transitions between winter and summer occur via states in the cluster C4, with a SST spatial distribution consisting on cold waters both in the north and in the south that clearly displays this transitional character.

Thus, we conclude that not only the SST spatial mean of each configuration was identified by the algorithm as characteristic of each cluster, but also other features in the spatial SST distribution are different at different seasons, being properly detected and classified by this K-means method.

%


\subsubsection{Self-Organizing Maps for SST}

\begin{figure}[t]
    \centering
    \includegraphics[width=0.5\textwidth]{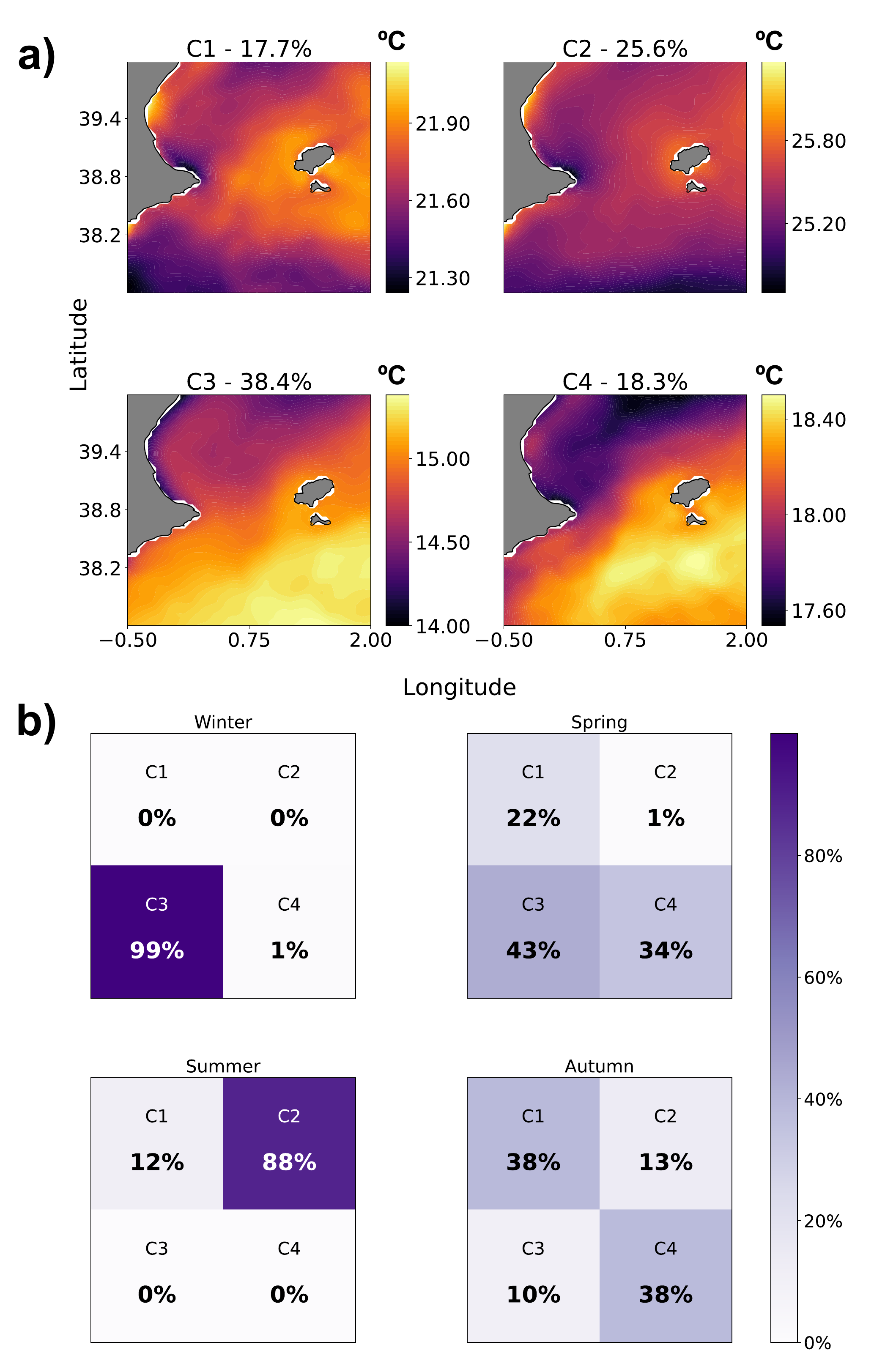}
    \caption{Panel a) shows the codebook vectors of SOM neurons, representing the sea surface temperature profile for each of the four clusters. Panel b) shows the distribution of data samples assigned to these clusters across different seasons (winter, spring, summer, and autumn).}
    \label{fig:seasons_and_clusters_SOM_temp}
\end{figure}

We now apply SOM clustering to the Mediterranean SST daily configurations. As for K-means, we use four clusters, each represented by a neuron and characterized by a codebook vector, i.e. a vector of pixel weights that is representative of the SST spatial configurations classified into that cluster. But now the relative location of the neurons is relevant and specified beforehand, here in two dimensions. Specifically we locate the four neurons in a 2x2 square grid, so that there is a distance of 1 between vertically or horizontally separated neurons, and $\sqrt{2}$ between the ones across the diagonals. Thus, these last neurons will encode clusters more different between them than the ones horizontally or vertically separated. Neurons, and then the clusters, are labelled as C1, C2 - the ones in the upper row - and C3 and C4 - the ones in the lower row.

As for the K-means method, the representative codebook vector of each cluster (Fig. \ref{fig:seasons_and_clusters_SOM_temp}a) has a distinct mean temperature, ordered from low to high as C3, C4, C1 and C2.  The coincidence of this ordering with the K-means results is not casual, but it is a consequence of our choosing in labelling the K-mean clusters according to the SOM cluster showing the largest similarity, as measured by the Jaccard analysis that will be described in Sect. \ref{subsec:resultsjaccard}. The similarities of K-means and SOM clustering go beyond the global mean temperature: There are clear similarities between the representative patterns (centroids and codebook vectors) shown in Figs. \ref{fig:seasons_and_clusters_kmeans_temp} and \ref{fig:seasons_and_clusters_SOM_temp}. In both cases, the C3 clusters show a clear southeast-northwest SST gradient, and C2 is more homogeneous, but with a colder region in the southern part. The transitional clusters C1 and C4 exhibit a structural similarity between the K-means centroids and SOM codebooks. Both methods consistently capture the spatial gradients of these intermediate states, with C4 representing the cooler transition and C1 the warmer one. This confirms that the identified transitional patterns are robust features of the dataset, independent of the specific clustering algorithm used. 

The seasonal dominance of the different clusters, shown in Fig. \ref{fig:seasons_and_clusters_SOM_temp}b) is very similar to the K-means case, with a dominance of winter configurations in C3, which also has significant presence in spring, and dominance of summer configurations in C2, with the transitional C1 and C4 clusters well represented in autumn and spring. As expected, the most distant states, C3 and C2, corresponding to winter and summer, appear located across the largest distance in the SOM grid, i.e. across the diagonal.

\subsection{Kinetic Energy}
\label{subsec:resultsKE}

This section shifts our focus to the spatial distribution of KE.
We utilize three distinct clustering methodologies:
K-means, SOM, and InfoMap.
Consistent with the SST analysis, where silhouette scores
(Figure \ref{fig:grid_size_vs_silhouette}) suggested
four clusters as optimal for K-means and SOM,
we also impose four KE clusters.

\subsubsection{K-means for KE}

\begin{figure}[t]
    \centering
    \includegraphics[width=8.3cm]{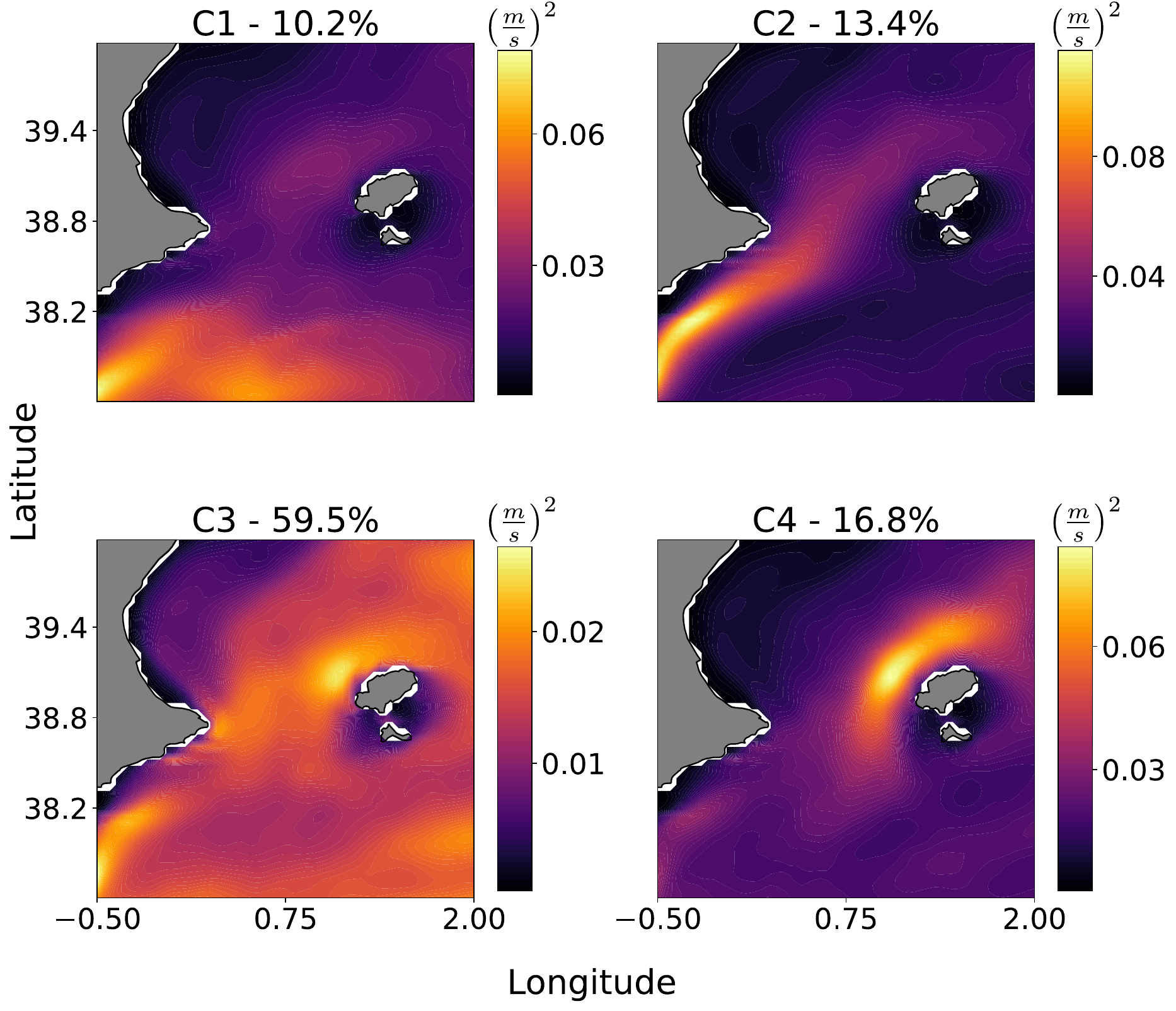}
    \caption{Centroids of the four clusters obtained from the application of the K-means clustering algorithm on the KE field. Each panel shows the average spatial distribution of KE in each cluster, with the percentage indicating the proportion of total samples assigned to that cluster. The color bar indicates the magnitude of KE.}
    \label{fig:kinetic_energy_kmeans}
\end{figure}

The centroids of the four clusters obtained with K-means are shown in Figure \ref{fig:kinetic_energy_kmeans}. They provide insight into the dominant KE regimes and their spatial characteristics. As in the SST case, the labels of the clusters are assigned according to their larger similarity to the SOM clusters, for which these labels have a topological significance.

C4, with 16.8\% of the data samples,
indicates a region of high KE north of Ibiza Island,
suggesting a localized and strong feature, reminiscent of the Balearic Current associated to the Balearic front and arising as a northward deflection of the Northern Current \citep{Millot1999circulation,Barral2021characterization}.
In contrast C3, which represents 59.5\% of the observations,
exhibits the  lowest energy intensity
and a more homogeneous distribution. This suggests that
the majority of the entries fall within a
regime characterized by low and uniform KE,
possibly corresponding to periods with weaker currents.
C2, with 13.4\% of the data, reveals a
medium-intensity energy feature
along the southern coast of the Iberian Peninsula.
Finally, C1, representing 10.2\% of the data,
shows a spatial distribution of KE that appears to be associated with a more diffuse
feature in the southern part of the domain, possibly less persistent
than those of C2 and C4, which may be associated to the Algerian Current. We notice that the Balearic Current appears in all centroids as a local maximum in KE, although only in C4 it is the most visible and dominant feature.

It is important to note that the centroids in K-means
tend to simplify complex patterns
by focusing on the mean energy distribution
within each cluster. This is effective for
identifying dominant regimes, but it will smooth out
small-scale features.

\subsubsection{Self-Organizing Maps for KE}

\begin{figure}[t]
    \centering
    \includegraphics[width=8.3cm]{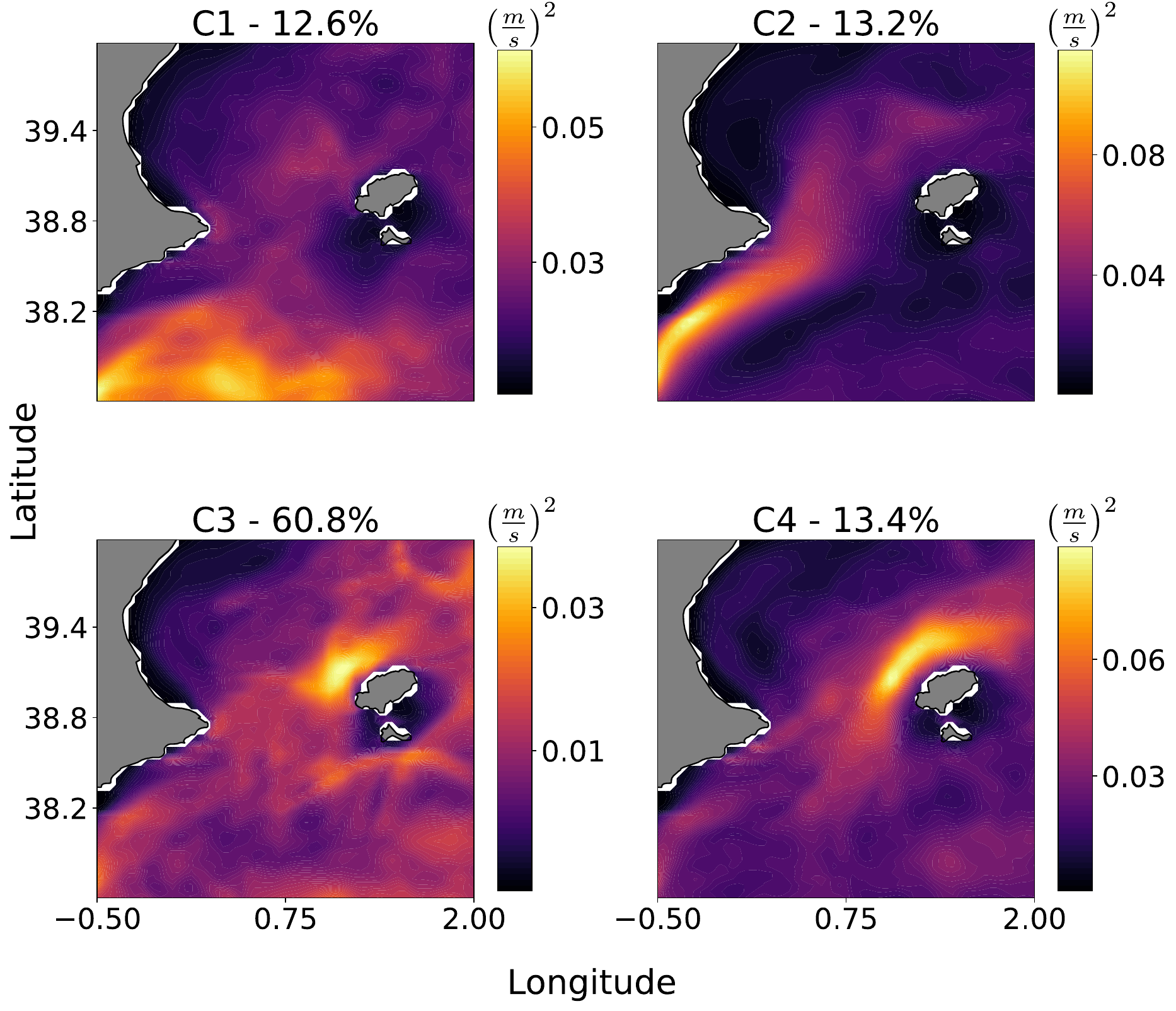}
    \caption{Codebook vectors of the four neurons encoding the representative spatial state of four clusters of KE resulting from the application of the SOM clustering method. The color bar indicates the magnitude of KE.}
    \label{fig:kinetic_energy_SOM}
\end{figure}

Figure \ref{fig:kinetic_energy_SOM} presents the results of the SOM analysis applied to the KE data. A comparison with the K-means results (Fig. \ref{fig:kinetic_energy_kmeans}) reveals a significant convergence between the two methods, both in the spatial topology of the patterns and in their statistical prevalence. This consistency confirms that the identified energetic regimes are robust physical modes of the region rather than artifacts of a specific algorithm.

As with K-means, the most dominant regime in SOM is captured by cluster C3 (60.8\%), which virtually mirrors the K-means background cluster C3 (59.5\%). Both represent a relatively calm state with low energy levels distributed across the domain, characteristic of periods without a very intense mesoscale activity.

The remaining high-energy configurations are distributed among three clusters that align closely with their K-means counterparts. Cluster C4 (13.4\%) corresponds to the K-means C4 (16.8\%) and clearly identifies the elevated kinetic energy of the \textit{Balearic Current} north of Ibiza Island. Similarly, Cluster C2 (13.2\%) matches the K-means C2 (13.4\%), capturing the energy feature along the southern coast of the Iberian Peninsula. Finally, Cluster C1 (12.6\%) parallels the K-means C1 (10.2\%), representing the energetic intrusions in the southern part of the domain associated with the \textit{Algerian Current}. The ability of both SOM and K-means to independently isolate these three specific high-energy modes, despite the differences in their underlying mathematical principles,
strongly suggests that the circulation in the Ibiza Channel oscillates between these well-defined dynamical states.

\subsubsection{InfoMap for KE}

\begin{figure}[t]
    \centering
    \includegraphics[width=8.3cm]{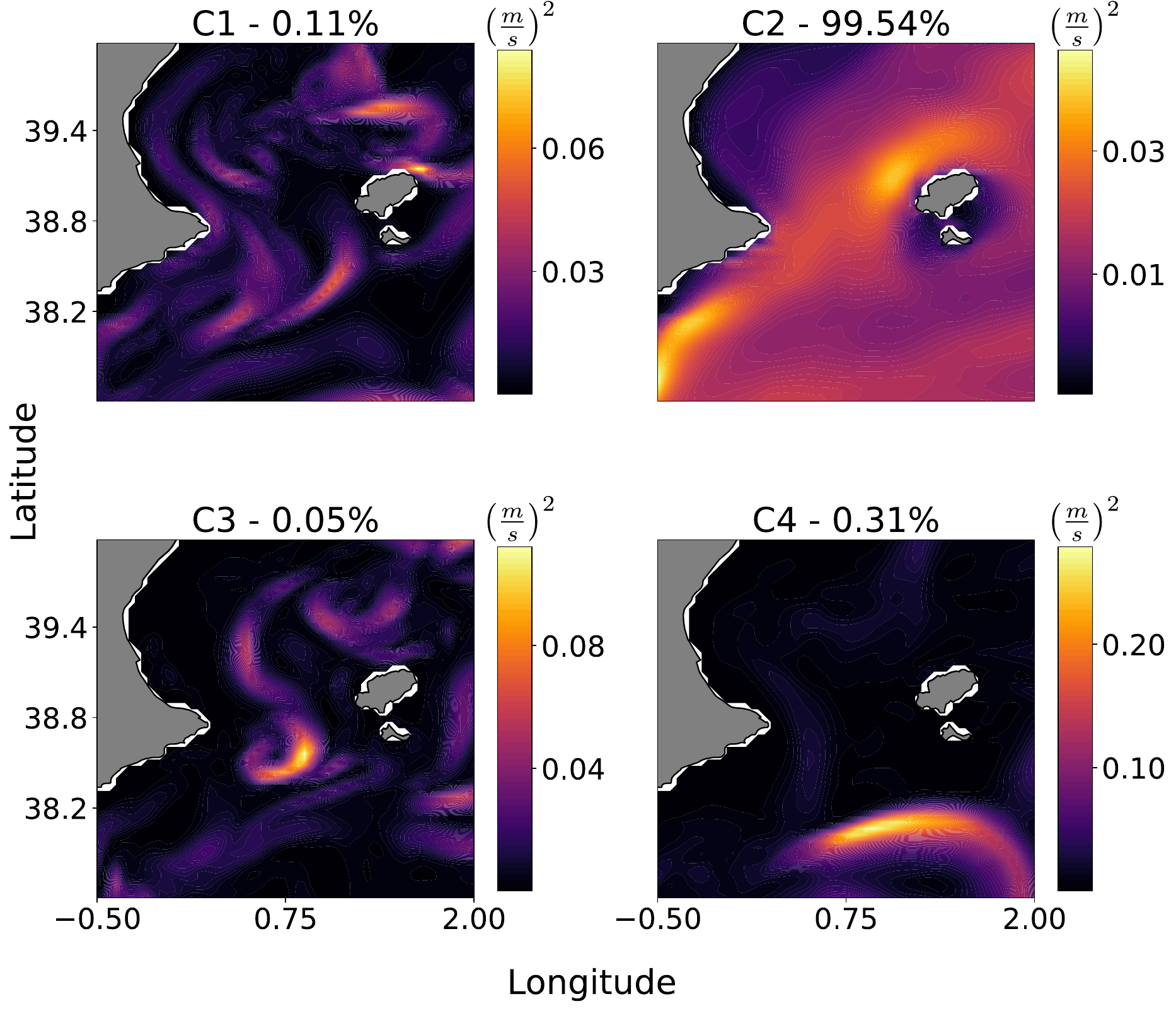}
    \caption{Spatial distribution of KE centroids for clusters identified using the InfoMap community detection method. Each panel shows the mean KE pattern of the populations classified in each cluster (C1-C4). Color intensity indicates KE magnitude}
    \label{fig:kinetic_energy_infomap}
\end{figure}

The centroids of the clusters of KE configurations obtained from InfoMap are shown in Fig. \ref{fig:kinetic_energy_infomap}. The threshold used to select only sufficiently high configuration similarities (mean cosinus similarity plus 1.27 times the standard deviation of all similarity values, see Sect. \ref{subsec:infomap}), leads, for the present dataset, to a partition consisting of four clusters. The labelling of the clusters is arbitrary.  It is known that InfoMap is not significantly affected by the so called \textit{resolution limit} \citep{fortunato2007resolution,Kawamoto2015estimating},  meaning that it can detect clusters composed by very few configurations. Indeed, nearly all configurations are assigned to cluster C2, containing 99,54\% of them. The centroid of this cluster is relatively homogeneous, with low energy levels, and some maxima on the Balearic Current and in a jet flowing northwards from the southern Mediterranean Spanish coast, indicating that the cluster contains KE configurations with these structures. Note the similarity of this centroid with that of C3 from the K-means (Fig. \ref{fig:kinetic_energy_kmeans}) and SOM (Fig. \ref{fig:kinetic_energy_SOM}) methods (which is also the most populated in these methods), and the low-energy levels of this cluster in the three methods. The centroids of InfoMap's clusters C1, C3 and C4 are much more structured, with features indicating the presence of eddies, jets, and other mesoscale structures. InfoMap assigns very few configurations to these clusters, so that despite the average nature of centroids, small-scale features are not smoothed out and remain visible. In particular, the centroid in C4 displays a high-energy jet in the region of the Algerian Currrent and the Algerian eddies \citep{sayol2013comparative}. 

To further understand the relevance of this 'intense south jet' energy configuration we applied K-means and SOM algorithms to subsets of the whole data set. We find that these two methods clearly identified a distinct cluster showing this feature when applied to the first thousand days of the dataset. This confirms that the configurations clustered by InfoMap as C4 represent real, physically consistent high-energy events rather than an artifact, despite comprising only 0.31\% of the total data. K-means and SOM extract only the most dominant configurations, thus hiding this 'south jet' configuration behind the other, most prevalent, energy distributions found. The discrepancy between this clear isolation in InfoMap and its dilution in the full K-means and SOM results highlights the different sensitivities of the algorithms. While K-means and SOM tend to average out such rare, low-populated configurations into broader dominant regimes when applied to the full 17-year dataset, InfoMap effectively acts as an anomaly detector, unveiling these fine-scale and localized dynamics. Similarly, InfoMap's C1 (0.11\%) and C3 (0.05\%) clusters identify medium-intensity eddies with spatial patterns suggesting localized rotational features (on the order of tens of kilometers) that were not apparent in the global results of the other methods. This underlines the complementary nature of the different clustering methodologies used in this paper, where InfoMap proves essential for comprehensively analyzing the rare, high-energy tails of the oceanographic distribution. A further comparison is performed in the next section.

\subsection{Cluster-analysis comparative for SST and KE: Jaccard index}
\label{subsec:resultsjaccard}

\begin{figure}[t]
    \centering
    \includegraphics[width=8.3cm]{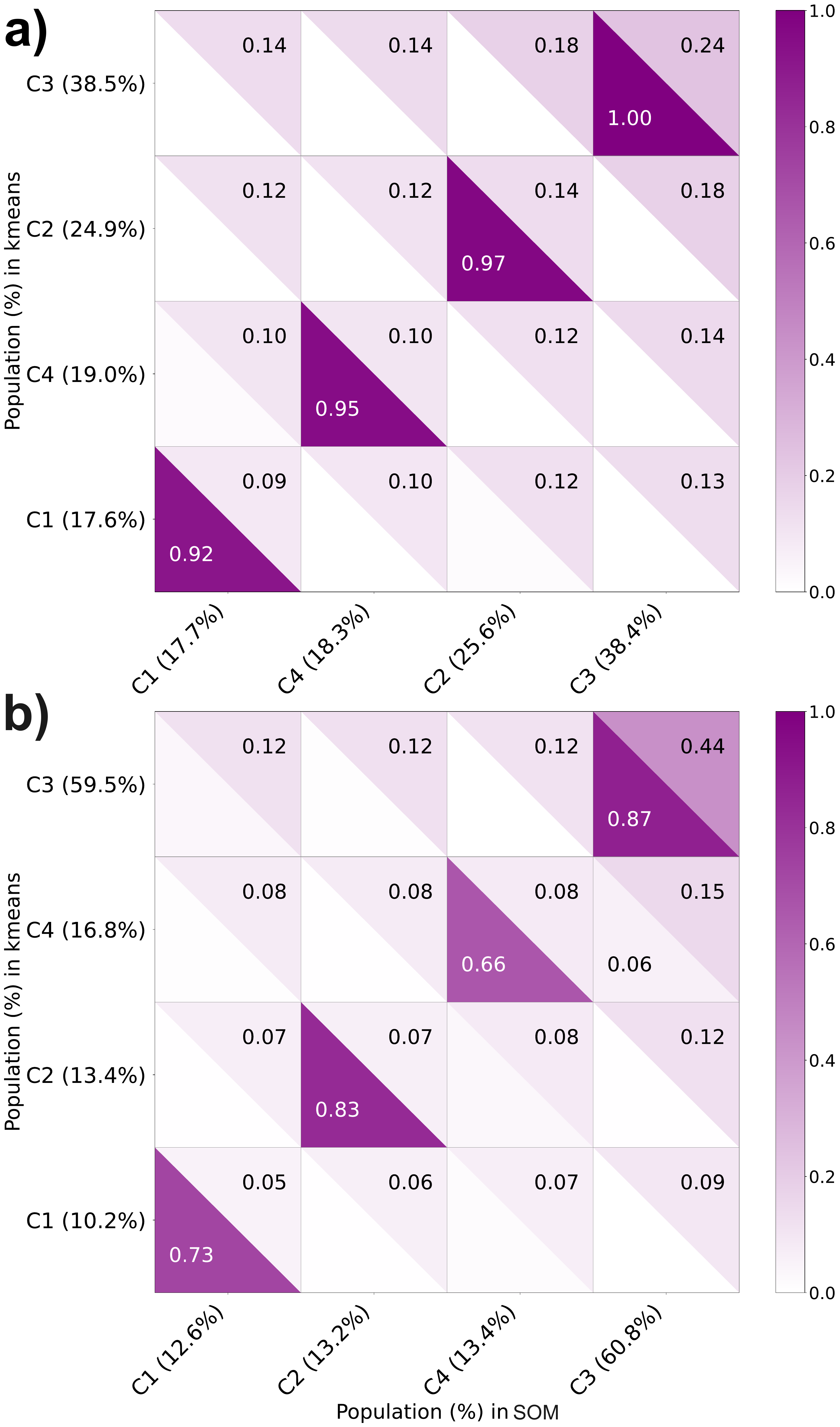}
    \caption{Jaccard similarity matrices comparing SOM and K-means clustering for (a) Sea Surface Temperature (SST) and (b) Kinetic Energy (KE). Matrix values range from 0.0 to 1.0, where darker purple indicates higher similarity (overlap). In each cell, the lower-left triangle represents the actual Jaccard index between the two methods, while the upper-right triangle represents a baseline Jaccard index calculated using randomized samples that preserve the original cluster sizes of both the K-means and SOM populations. The axes indicate the population percentage for each cluster}
    \label{fig:combined_jaccard_similarity}
\end{figure}

The Jaccard similarity index matrices comparing the clustering results from K-means and SOM are presented in Figure \ref{fig:combined_jaccard_similarity} for both SST and KE data. In the upper panel (SST), we observe an exceptionally high degree of similarity between the clusters generated by SOM and K-means. The diagonal elements exhibit Jaccard indices close to unity, indicated by the deep purple shades. This near-perfect correspondence confirms that the two methods provide an identical and robust classification of SST configurations into the four seasonal regimes.

The lower panel (KE) also reveals a strong consistency between the methods, contrasting with the potential divergence often expected when clustering complex, high-variability fields. While the Jaccard indices are slightly lower than those for SST, reflecting the higher complexity of the kinetic energy field, the diagonal dominance is clear, with values ranging from 0.66 to 0.87. This indicates that the dynamical regimes identified by SOM (the background state, the Balearic Current, the Algerian Current, and the anticyclonic eddy field) are robustly matched by the K-means centroids. A  slightly lower agreement (0.66) is found in the cluster associated with the localized eddy activity (C4 in K-means, 16.8\%; C4 in SOM, 13.4\%), suggesting that while both methods identify this high-energy feature, the exact boundaries of the partition differ slightly due to the topological constraints of SOM versus the variance-minimization of K-means. Nevertheless, the overall picture is one of convergence, validating that the identified patterns are intrinsic to the physical system rather than artifacts of a specific algorithm.

Another observation from the figure is the consistently higher Jaccard indices in the lower-left triangles (representing the actual method comparison) compared to the upper-right triangles. The latter represent the indices obtained by comparing the K-means clusters against randomized clusters that maintain the exact population sizes of the SOM results. The fact that the actual similarity scores exceed the randomized baseline in both panels confirms that the detected clusters, in both SST and KE, reflect  statistically significant patterns within the data and are not merely due to chance.

\section{Conclusions}
\label{sec:conclusions}

In this paper, we addressed a comparative analysis of clustering methods for SST and KE data of the Western Mediterranean, restricting our focus to surface data at  both large-scale and mesoscale dynamics.

For SST, the analysis showed that both K-means and SOM partitioned the data into four distinct regimes—a number suggested a priori by silhouette-score analysis. These regimes were highly consistent between the two methods and strongly aligned with the seasonal cycle. Beyond their temporal synchronization, the clusters captured distinct physical surface patterns consistent with the literature \citep{DOrtenzio2005}:
\begin{itemize}
\item The \textbf{Winter regime} (C3) displays a marked northwest-southeast thermal gradient, separating the colder northern waters from the warmer southern basin.
\item The \textbf{Summer regime} (C2) is characterized by high mean temperatures and a relatively homogeneous spatial distribution, yet it distinctly preserves a colder signature along the southern boundary, corresponding to the surface inflow of Atlantic waters.
\item The \textbf{Transitional regimes} (C1 and C4) capture the seasonal inversion of these gradients, identifying the intermediate states where the north-south temperature difference re-adjusts.
\end{itemize}
Furthermore, the analysis of the temporal evolution of these partitions uncovered a shortening of the winter SST regime over the study period, suggesting a dynamic response of the surface layer to long-term climate variability.

In the analysis of Kinetic Energy, we found a striking convergence between K-means and SOM. Quantitative comparison using the Jaccard index confirmed strong agreement, with both methods successful at isolating the dominant hydrodynamic modes of the Balearic Sea. These identified clusters correspond to well-known circulation features described in the literature \citep{Millot1999circulation, Barral2021characterization}:
\begin{itemize}
\item A \textbf{Low-energy Background State} (C3, $\sim$60\% of data): A relatively low-energy regime representing the typical conditions when the most intense mesoscale activity is absent.
\item The \textbf{Balearic Current} (C4): identified as a localized and strong feature north of Ibiza Island, arising as a northward deflection of the Northern Current.
\item The \textbf{Algerian Basin Variability} (C2 and C4): High-energy features identified along the southern coast of the Iberian Peninsula and in the southern part of the domain, associated with the instabilities of the Algerian Current and the presence of mesoscale eddies.
\end{itemize}

The application of InfoMap provided a complementary perspective. While K-means and SOM effectively classified the
 dominant dynamical regimes, InfoMap operated essentially as an anomaly detector. It successfully isolated rare, high-magnitude events, specifically the most intense realizations of the southern jet (comprising only 0.3\% of the data),
 which were statistically merged into broader high-energy clusters by the partition-based methods.

Our study suggests that there is no single "best" algorithm, but rather a hierarchy of utility for oceanographic data mining. K-means and SOM provide a reliable, robust framework for identifying dominant climatological modes and their transitions, while InfoMap offers the unique ability to unveil fine-scale, extreme events that lie in the tails of the probability distribution. We hope that this quantitative framework for characterizing Mediterranean surface dynamics holds significant implications for the application of Artificial Intelligence tools in marine ecosystem management, operational oceanography, and regional climate research.

\section*{Code and Data Availability}

The Mediterranean Sea Physical Reanalysis product (MEDSEA\_MULTIYEAR\_PHY\_006\_004) used in this study is available from the Copernicus Marine Service (CMEMS) at \url{https://doi.org/10.25423/CMCC/MEDSEA_MULTIYEAR_PHY_006_004_E3R1} \citep{Escudier2020_dataset}. The subsets of this product that have been analyzed here, as well as the Python code implementing the procedures, are accessible in \citet{DataCode}, \url{https://doi.org/10.20350/digitalCSIC/18346}. 

\authorcontribution{V.R.-M., E.S.-G., C.L., J.J.R. and E.H.-G (all the authors) conceived the method, V.R.-M. conducted the numerical data analysis, V.R.-M., E.S.-G., C.L., J.J.R. and E.H.-G (all the authors) analyzed the results, contributed in the writing, and revised the manuscript.}

\competinginterests{The authors declare that they have no competing interests.}

\begin{acknowledgements}
This work was partially supported by project PID2021-123352OB-C32  funded by MICIU/AEI/10.13039/501100011033  and FEDER, "Una manera de hacer Europa". E.S.-G., J.J.R. and E.H.-G acknowledge support by the Mar\'{\i}a de Maeztu project CEX2021-001164-M funded by MICIU/AEI/10.13039/501100011033.
\end{acknowledgements}

\bibliographystyle{copernicus}
\bibliography{references.bib}

\end{document}